\documentclass[a4paper,11pt]{article}
\usepackage[pdftex]{graphicx}
\usepackage[cmex10]{amsmath}
\usepackage{amssymb}
\usepackage[margin=1.2in]{geometry}
\usepackage[hyphens]{url}

\usepackage{hyperref}

\usepackage{breakurl}
\usepackage{algorithm}
\usepackage[noend]{algpseudocode}
\usepackage{amsthm}
\usepackage{lipsum}

\newcommand{\atan}{\ensuremath{\operatorname{atan}}}

\begin{document}

\title{Efficient implementation of elementary functions in the medium-precision range}
\author{Fredrik Johansson\footnote{INRIA Bordeaux} \footnote{fredrik.johansson@gmail.com}}
\date{}

\maketitle

\begin{abstract}
We describe a new implementation of the elementary transcendental
functions exp, sin, cos, log and atan for variable precision up to
approximately 4096 bits. Compared to the MPFR library,
we achieve a maximum speedup ranging
from a factor 3 for cos to 30 for atan.
Our implementation uses table-based
argument reduction together with rectangular splitting to evaluate
Taylor series. We collect denominators to reduce the number of
divisions in the Taylor series, and avoid overhead by doing all
multiprecision arithmetic using the mpn layer of the GMP library.
Our implementation provides rigorous error bounds.
\end{abstract}


\section{Introduction}
Considerable effort has been made to optimize
computation of the elementary transcendental
functions in IEEE~754 double precision arithmetic (53 bits) subject
to various constraints
\cite{dedinechin:inria-00071446,daramy2003cr,dukhan2014methods,harrison1999computation,metalibm}.
Higher precision is indispensable for computer algebra and
is becoming increasingly important in scientific applications~\cite{bailey2012high}.
Many libraries have been developed for arbitrary-precision arithmetic.
The de facto standard is arguably MPFR~\cite{Fousse2007},
which guarantees correct rounding to any requested number of bits.

Unfortunately, there is a large performance gap between double precision
and arbitrary-precision libraries.
Some authors have helped bridge this gap by developing
fast implementations targeting
a fixed precision, such as 106, 113 or 212 bits
\cite{thall2006extended,6081400,hida2007library}. However, these
implementations generally do not provide rigorous error bounds
(a promising approach to remedy this situation is \cite{metalibm}),
and performance optimization in the range of several hundred bits still appears to be lacking.

The asymptotic difficulty of computing elementary functions is well understood.
From several thousand bits and up, the bit-burst algorithm or the
arithmetic-geometric mean algorithm coupled with Newton iteration effectively
reduce the problem to integer multiplication, which has
quasilinear complexity \cite{brent1976complexity,mca}. Although such high
precision has uses, most
applications beyond double precision only require modest
extra precision, say a few hundred bits or rarely a few thousand bits.

In this ``medium-precision'' range beyond double precision and up
to a few thousand bits, i.e.\ up to perhaps a hundred words
on a 32-bit or 64-bit computer, there are two principal hurdles
in the way of efficiency. First, the cost of
$(n \times n)$-word multiplication or division grows
quadratically with $n$, or almost quadratically if Karatsuba
multiplication is used, so rather than
``reducing everything to multiplication'' (in the words of \cite{steelreduce}),
we want to do as little multiplying as possible. Secondly, since
multiprecision arithmetic currently has to be done in software, every
arithmetic operation potentially involves overhead for function calls,
temporary memory allocation, and case distinctions based on signs and
sizes of inputs; we want to avoid as much of this bookkeeping as possible.

In this work, we consider the
five elementary functions exp, sin, cos, log, atan of a real variable,
to which all other real and complex elementary functions can be delegated
via algebraic transformations.
Our algorithm for all five functions follows the well-known strategy of argument
reduction based on functional equations
and lookup tables as described in section~\ref{sect:argred},
followed by evaluation of Taylor series.
To keep overhead at a minimum,
all arithmetic uses
the low-level mpn layer of the GMP library~\cite{gmp},
as outlined in section~\ref{sect:fixed}.

We use lookup tables in arguably the simplest possible way,
storing values of the function itself on a regularly spaced grid.
At high precision, a good space-time tradeoff is achieved by using 
bipartite tables.
Several authors have studied the problem of constructing optimal
designs for elementary functions
in resource-constrained settings, where it is important to minimize
not only the size of the tables but also the numerical error
and the complexity of circuitry to implement the arithmetic operations
\cite{de2005multipartite}, \cite{schulte1999approximating}, \cite{stine1999symmetric}.
We ignore such design parameters since
guard bits and code size are cheap in our setting.

While implementations in double precision often use
minimax or Chebyshev polynomial approximations,
which require somewhat fewer terms than Taylor series
for equivalent accuracy, Taylor series
are superior at high precision
since the evaluation can be done faster.
Smith's rectangular splitting algorithm~\cite{Smith1989}
allows evaluating a degree-$N$ truncated Taylor series
of suitable type using $O(\sqrt{N})$
$(n \times n)$-word multiplications whereas
evaluating a degree-$N$ minimax polynomial using
Horner's rule requires $O(N)$ such multiplications.

The main contribution of the paper, described in
section~\ref{sect:taylor}, is an
improved version of Smith's rectangular splitting algorithm
for evaluating Taylor series, in which we use fixed-point
arithmetic efficiently and avoid most divisions.
Section~\ref{sect:toplevel} describes
the global algorithm including
error analysis.

Our implementation of the elementary functions
is part of version~2.4.0 of
the open source arbitrary-precision interval
software Arb~\cite{Johansson:2014:ACL:2576802.2576828}.
The source code can be retrieved from~\cite{fjarbsource}.

Since the goal is to do interval arithmetic, we
compute a rigorous bound for the numerical error.
Unlike MPFR, our code does not output a correctly rounded
floating-point value.
This more of a difference in the interface
than an inherent limitation of the algorithm,
and only accounts for a
small difference in performance (as explained in Section~\ref{sect:toplevel}).

Our benchmark results in section~\ref{sect:bench} show
a significant speedup compared to the current version (3.1.2) of MPFR.
MPFR uses several different algorithms depending on the precision and
function~\cite{mpfralg}, including Smith's algorithm in some cases.
The large improvement is in part due to our use of lookup tables (which MPFR does not use)
and in part due to the optimized Taylor series evaluation and elimination
of general overhead.
Our different elementary functions also have similar
performance to each other. Indeed, the algorithm is nearly the same for
all functions, which simplifies the software design and aids proving correctness.

While our implementation allows variable precision up to a few thousand bits,
it is competitive in the low end of the range with
the QD library~\cite{hida2007library}
which only targets 106 or 212 bits. QD uses a combination of lookup tables,
argument reduction, Taylor series, and Newton iteration for inverse functions.

\section{Fixed-point arithmetic}
\label{sect:fixed}

We base our multiprecision arithmetic
on the GMP library~\cite{gmp} (or the fork MPIR~\cite{mpir}),
which is widely available
and optimized for common CPU architectures.
We use the mpn layer of GMP, since the
mpz layer has unnecessary overhead.
On the mpn level, a multiprecision integer is an array of
limbs (words). We assume that a limb is either $B = 32$ or $B = 64$ bits,
holding a value between $0$ and $2^B-1$.
We represent a real number in fixed-point format
with $Bn$-bit precision using~$n$ fractional limbs and
zero or more integral limbs.
An $n$-limb array thus represents a value in the range $[0,1-\text{ulp}]$,
and an $(n+1)$-limb array represents a value in the range $[0,2^B-\text{ulp}]$
where $\text{ulp} = 2^{-Bn}$.

An advantage of fixed-point over floating-point arithmetic
is that we can add numbers without any rounding
or shift adjustments.
The most important GMP functions
are shown in Table \ref{tab:fixedpoint}, where
$X, Y, Z$ denote fixed-point numbers with the same number of limbs
and $c$ denotes a single-limb unsigned integer. Since the first five
functions return carry-out or borrow, we can also use them
when $X$ has one more limb than $Y$.

\begin{table}[ht!]
\centering
\caption{Fixed-point operations using GMP.}
\begin{tabular}{ l l }
\texttt{mpn\_add\_n} & $X \gets X + Y$ (or $X \gets Y + Z$) \\
\texttt{mpn\_sub\_n} & $X \gets X - Y$ (or $X \gets Y - Z$) \\
\texttt{mpn\_mul\_1} & $X \gets Y \times c$ \\
\texttt{mpn\_addmul\_1} & $X \gets X + Y \times c$ \\
\texttt{mpn\_submul\_1} & $X \gets X - Y \times c$ \\
\texttt{mpn\_mul\_n} & $X \gets Y \times Z$ \\
\texttt{mpn\_sqr} & $X \gets Y \times Y$ \\
\texttt{mpn\_divrem\_1} & $X \gets Y / c$ \\
\end{tabular}
\label{tab:fixedpoint}
\end{table}

The first five GMP functions in Table~\ref{tab:fixedpoint}
are usually implemented in
assembly code, and we therefore try to push the
work onto those primitives.
Note that multiplying two $n$-limb fixed-point numbers involves computing
the full $2n$-limb product and throwing away the~$n$ least significant
limbs. We can often avoid explicitly copying the high limbs
by simply moving the pointer into the array.

The mpn representation does not admit negative numbers. However, we
can store negative numbers implicitly using two's complement representation
as long as we only add and subtract fixed-point numbers with the same number of
limbs. We must then take care to ensure that the value is positive before
multiplying or dividing.

We compute bounds for all errors when doing fixed-point arithmetic.
For example, if $X$ and $Y$ are fixed-point
numbers with respective errors $\varepsilon_1$, $\varepsilon_2$,
then their sum has error bounded by $|\varepsilon_1| + |\varepsilon_2|$,
and their product, rounded to a fixed-point number using a single truncation,
has error bounded by
$$|Y| |\varepsilon_1| + |X| |\varepsilon_2|
    + |\varepsilon_1 \varepsilon_2| + (1~\text{ulp}).$$
If $c$ is an exact integer, then the product
$X \times c$ has error bounded by $|\varepsilon_1||c|$, and
the quotient $X/c$, rounded to a fixed-point number using a single truncation,
has error bounded by $|\varepsilon_1|/|c| + (1~\text{ulp})$.
Similar bounds are used for other operations that arise in the implementation.

In parts of the code, we use a single-limb variable
to track a running error bound measured in ulps,
instead of determining
a formula that bounds the cumulative error in advance.
This is convenient, and cheap compared to
the actual work done in the multiprecision arithmetic operations.

\section{Argument reduction}
\label{sect:argred}

The standard method to evaluate elementary functions
begins with one or several argument reductions to restrict the input to
a small standard domain. The function is then
computed on the standard domain, typically using a polynomial approximation
such as a truncated Taylor series,
and the argument reduction steps are inverted to recover
the function value \cite{mca}, \cite{muller2006elementary}.

As an example, consider the exponential function $\exp(x)$.
Setting $m = \lfloor x / \log(2) \rfloor$ and $t = x - m \log(2)$,
we reduce the problem to computing $\exp(x) = \exp(t) 2^m$ where~$t$
lies in the standard domain $[0, \log(2))$.
Writing $\exp(t) = [\exp(t/2^r)]^{2^r}$, we can further reduce the argument
to the range $[0, 2^{-r})$ at the expense of $r$ squarings,
thereby improving the rate of convergence of the Taylor series.
Analogously, we can reduce to the intervals $[0,\pi/4)$ for
sin and cos, $[0,1)$ for atan, and $[1,2)$ for log,
and follow up with~$r$ further transformations
to reduce the argument to an interval of width $2^{-r}$.

This strategy does not require precomputations
(except perhaps for the constants $\pi$ and $\log(2)$), and is
commonly used in arbitrary-precision libraries
such as MPFR \cite{mpfralg}.

The argument reduction steps can be accelerated using lookup tables.
If we precompute $\exp(i/2^r)$ for $i = 0 \ldots 2^r-1$,
we can write $\exp(x) = \exp(x-i/2^r) \exp(i/2^r)$ where
$i = \lfloor 2^r x \rfloor$. This achieves $r$ halvings worth
of argument reduction for the cost of just a single multiplication.
To save space, we
can use a bipartite (or multipartite) table, e.g.\
writing $\exp(x) = \exp(x-i/2^r-j/2^{2r}) \exp(i/2^r) \exp(j/2^{2r})$.

This recipe works for all elementary functions.
We use the following formulas, in which $x \in [0, 1)$,
$q = 2^r$, $i = \lfloor 2^r x \rfloor$, $t = i/q$, $w = x-i/q$,
$w_1 = (qx-i)/(i+q)$, and $w_2 = (qx-i)/(ix+q)$:
\begin{align*}
\exp(x) & = \exp(t) \exp(w) \\
\sin(x) & = \sin(t) \cos(w) + \cos(t) \sin(w) \\
\cos(x) & = \cos(t) \cos(w) - \sin(t) \sin(w) \\
\log(1+x) & = \log(1+t) + \log(1+w_1) \\
\atan(x) & = \atan(t) + \atan(w_2)
\end{align*}

The sine and cosine are best computed simultaneously. The argument reduction formula for the logarithm is cheaper than for the other functions, since it requires $(n \times 1)$-word operations and no $(n \times n)$-word multiplications or divisions. The advantage of using lookup tables is greater for log and atan than for exp, sin and cos, since the ``argument-halving'' formulas for log and atan involve square roots.

If we want $p$-bit precision and chain together $m$ lookup tables worth~$r$ halvings each, the total amount of space is $m p 2^r$ bits, and the number of terms in the Taylor series that we have to sum is of the order $p / (r m)$. Taking~$r$ between~4 and~10 and $m$ between~1 and~3 gives a good space-time tradeoff. At lower precision, a smaller $m$ is better.

\begin{table}[ht!]
\centering
\caption{Size of lookup tables.}
\begin{tabular}{ l l l l r l }
Function & Precision & $m$ & $r$ & Entries &  Size (KiB) \\ \hline
exp & $\le 512$     &     1  & 8   &   178 &    11.125 \\
exp & $\le 4608$    &     2  & 5   & 23+32  &   30.9375 \\
sin & $\le 512$     &     1  & 8   &   203 &    12.6875 \\
sin & $\le 4608$    &     2  & 5   & 26+32 &    32.625 \\
cos & $\le 512$     &     1  & 8   &   203 &    12.6875 \\
cos & $\le 4608$    &     2  & 5   & 26+32 &    32.625 \\
log & $\le 512$     &     2  & 7   & 128+128  &   16 \\
log & $\le 4608$    &     2  & 5   & 32+32  &   36 \\
atan & $\le 512$    &     1  & 8  &    256  &   16 \\
atan & $\le 4608$   &     2  & 5  &  32+32  &   36 \\ \hline
Total         &     &    &       &   &   236.6875
\end{tabular}
\label{tab:tablesize}
\end{table}

Our implementation uses the table parameters
shown in Table~\ref{tab:tablesize}. For each function, we use a
fast table up to 512 bits
and a more economical table from 513 to 4608 bits,
supporting function evaluation at precisions just beyond
4096 bits plus guard bits.
Some of the tables have
less than $2^r$ entries since they end near $\log(2)$
or $\pi/4$.
A few more kilobytes are used to store precomputed values
of $\pi/4$, $\log(2)$, and coefficients of Taylor series.

The parameters in Table~\ref{tab:tablesize} were chosen based on experiment
to give good performance at all precisions
while keeping the total size (less than 256~KiB)
insignificant compared to the overall space requirements of most applications
and small enough to fit in a typical L2 cache.
For simplicity, our code uses static precomputed tables,
which are tested against MPFR to verify that
all entries are correctly rounded.

The restriction to 4096-bit and lower precision is done since
lookup tables give diminishing returns at higher precision
compared to asymptotically fast algorithms
that avoid precomputations entirely.
In a software implementation, there is no practical
upper limit to the size of lookup tables that can be used.
One could gain efficiency by using auxiliary code to
dynamically generate tables that are optimal for a given application.

\section{Taylor series evaluation}
\label{sect:taylor}

After argument reduction, we need
to evaluate a truncated Taylor series, where we
are given a fixed-point argument $0 \le X \ll 1$ and
the number of terms $N$ to add.
In this section, we present an algorithm that solves the problem
efficiently, with a bound for the rounding error.
The initial argument reduction
restricts the possible range of $N$,
which simplifies the analysis. Indeed,
for an internal precision of $p \le 4608$ bits and the 
parameters of Table~\ref{tab:tablesize}, 
$N < 300$ always suffices.

We use a version of Smith's algorithm to avoid expensive
multiplications~\cite{Smith1989}.
The method is best explained by an example. To evaluate
$$\atan(x) \approx x \sum_{k=0}^{N-1} \frac{(-1)^k t^k}{2k+1}, \quad t = x^2$$
with
$N = 16$, we pick the splitting parameter $m = \sqrt{N} = 4$
and write $\atan(x) /x \approx$
\begin{equation*}
\begingroup
\renewcommand*{\arraystretch}{1.3}
\begin{matrix}
  & [1              & - & \tfrac{1}{3}  t & + & \tfrac{1}{5}  t^2 & - & \tfrac{1}{7} t^3] & \,   \\
+ & [\tfrac{1}{9}  & - & \tfrac{1}{11} t & + & \tfrac{1}{13} t^2 & - & \tfrac{1}{15} t^3] & t^4 \\
+ & [\tfrac{1}{17} & - & \tfrac{1}{19} t & + & \tfrac{1}{21} t^2 & - & \tfrac{1}{23} t^3] & t^8 \\
+ & [\tfrac{1}{25} & - & \tfrac{1}{27} t & + & \tfrac{1}{29} t^2 & - & \tfrac{1}{31} t^3] & t^{12}.
\end{matrix}
\endgroup
\end{equation*}
Since the powers $t^2, \ldots, t^m$ can be recycled for each row,
we only need $2 \sqrt{N}$ full $(n \times n)$-limb
multiplications, plus $O(N)$ ``scalar'' operations, i.e.\
additions and $(n \times 1)$-limb divisions.
This ``rectangular'' splitting arrangement of the terms
is actually a transposition of Smith's ``modular'' algorithm,
and appears to be superior since Horner's rule
can be used for the outer polynomial evaluation
with respect to $t^m$ (see \cite{mca}).

A drawback of Smith's algorithm is that an $(n \times 1)$
division has high overhead compared to an
$(n \times 1)$ multiplication, or even an $(n \times n)$ multiplication
if~$n$ is very small.
In \cite{Johansson2014}, a different rectangular splitting
algorithm was proposed that uses $(n \times O(\sqrt{N}))$-limb
multiplications instead of scalar divisions, and also works in
the more general setting of holonomic functions. Initial experiments
done by the author suggest that the method of~\cite{Johansson2014} can
be more efficient at modest precision.
However, we found that another variation turns
out to be superior for the Taylor series of the elementary functions,
namely to simply collect several consecutive denominators in a single word,
replacing most $(n \times 1)$-word divisions
by cheaper $(n \times 1)$-word multiplications.

We precompute tables
of integers $u_k, v_k < 2^B$
such that
$1/(2k+1) = u_k / v_k$
and $v_k $ is the least common multiple
of $2i-1$ for several consecutive $i$ near $k$. To generate the
table, we iterate upwards from $k = 0$, picking
the longest possible sequence of terms on a common denominator without
overflowing a limb, starting a new subsequence from
each point where overflow occurs.
This does not necessarily give the least possible number
of distinct denominators, but it is close to optimal
(on average, $v_k$ is 28 bits wide
on a 32-bit system and 61 bits wide on a 64-bit system for $k < 300$).
The $k$ such that $v_k \ne v_{k+1}$
are
$$12, 18, 24, 29, \ldots, 226, 229, \ldots (\text{32-bit})$$ 
and
$$23, 35, 46, 56, \ldots, 225, 232, \ldots (\text{64-bit}).$$
In the supported range, we need at most one division
every three terms (32-bit) or every seven terms (64-bit), and
less than this for very small $N$.

We compute the sum backwards.
Suppose that the current partial sum is ${S / v_{k+1}}$.
To add $u_k / v_k$ when $v_k \ne v_{k+1}$,
we first change denominators by computing
$S \gets (S \times v_{k+1}) / v_k$.
This requires one $((n+1) \times 1)$
multiplication and one $((n+2) \times 1)$ division.
A complication arises if $S$ is a two's complemented negative value when we change
denominators, however in this case we can just ``add and subtract 1'',
i.e.\ compute
$$((S + v_{k+1}) \times v_k) / v_{k+1} - v_k$$
which costs
only two extra single-limb additions.

\begin{algorithm}
  \caption{Evaluation of the atan Taylor series}
  \label{alg:atan}
  \begin{algorithmic}[1]
    \Require $0 \le X \le 2^{-4}$ as an $n$-limb fixed-point number, $2 < N < 300$
    \Ensure $S \approx \sum_{k=0}^{N-1} \tfrac{(-1)^k}{2k+1} X^{2k+1}$ as an $n$-limb fixed-point number with $\le$ 2 ulp error
    \State $m \gets 2 \lceil \sqrt{N}/2\rceil$
    \State $T_1 \gets X \times X + \varepsilon$ \Comment{Compute powers of $X$, $n$ limbs each} \label{lst:line:xsquare1}
    \State $T_2 \gets T_1 \times T_1 + \varepsilon$
    \For{$(k = 4; \; k \le m; \; k \gets k + 2)$}
        \State $T_{k-1} \gets T_{k/2} \times T_{k/2-1} + \varepsilon$
        \State $T_k \gets T_{k/2} \times T_{k/2} + \varepsilon$
    \EndFor
    \State $S \gets 0$ \Comment{Fixed-point sum, with $n+1$ limbs}
    \For{$(k = N - 1; \; k \ge 0; \; k \gets k - 1)$}
        \If{$v_k \ne v_{k+1}$ \textbf{and} $k < N - 1$} \Comment{Change denominators}
            \If {$k$ is even}
                \State $S \gets S + v_{k+1}$  \label{lst:line:spluse}  \Comment{Single-limb addition}
            \EndIf
            \State $S \gets S \times v_k$ \Comment{$S$ temporarily has $n+2$ limbs} \label{lst:line:suse1}
            \State $S \gets S / v_{k+1} + \varepsilon$ \Comment{$S$ has $n+1$ limbs again}
            \If {$k$ is even}
                \State $S \gets S - v_k$  \label{lst:line:sminusd}  \Comment{Single-limb addition}
            \EndIf
        \EndIf
        \If{$k \bmod m = 0$}
            \State $S \gets S + (-1)^k u_k$  \label{lst:line:splusc1}  \Comment{Single-limb addition}
            \If{$k \ne 0$}
                \State $S \gets S \times T_m + \varepsilon$ \label{lst:line:suse2} \Comment{$((n+1) \times n)$-limb multiplication}
            \EndIf
        \Else
            \State $S \gets S + (-1)^k u_k \times T_{k \bmod m}$  \label{lst:line:splusc2} \Comment{Fused addmul of $n$ into $n+1$ limbs}
        \EndIf
    \EndFor
    \State $S \gets S / v_0 + \varepsilon$ \label{lst:line:suse3}
    \State $S \gets S \times X + \varepsilon$ \label{lst:line:finalmul}
    \State \Return $S$ \Comment{Only $n$ limbs}
  \end{algorithmic}
\end{algorithm}

Pseudocode for our implementation of the
atan Taylor series is shown in Algorithm~\ref{alg:atan}.
All uppercase variables denote fixed-point numbers, and all lowercase
variables denote integers.
We write $+ \varepsilon$ to signify a fixed-point operation that
adds up to 1~ulp of rounding error. All other operations are exact.

Algorithm~\ref{alg:atan} can be shown to be correct
by a short exhaustive computation.
We execute the algorithm symbolically for all allowed values of $N$.
In each step, we determine an upper bound for
the possible value of each fixed-point variable as well as its error,
proving that no overflow is possible
(note that $S$ may wraparound on lines \ref{lst:line:splusc1}
and \ref{lst:line:splusc2} since we use two's complement arithmetic
for negative values, and part of the proof is to verify that
$0 \le |S| \le 2^B - \text{ulp}$ necessarily holds before
executing lines \ref{lst:line:suse1},
\ref{lst:line:suse2}, \ref{lst:line:suse3}).
The computation proves that the error is bounded by 2 ulp
at the end.

It is not hard to see heuristically why the 2 ulp bound holds.
Since the sum is kept multiplied by a denominator which is close
to a full limb, we always have close to a full limb worth of guard bits.
Moreover, each multiplication by a power of $X$ removes
most of the accumulated error since $X \ll 1$.
At the same time, the numerators and denominators are
never so close to $2^B - 1$ that overflow is possible.
We stress that the proof depends on the particular content of the tables
$u$ and $v$.

Code to generate coefficients and prove correctness of Algorithm~\ref{alg:atan}
(and its variants for the other functions) is
included in the source repository~\cite{fjarbsource} in the form of a Python
script~{\tt verify\_taylor.py}.

Making small changes to Algorithm~\ref{alg:atan} allows
us to compute log, exp, sin and cos.
For log, we write
$\log(1+x) = 2 \operatorname{atanh}(x/(x+2))$,
since the Taylor series for atanh
has half as many nonzero terms.
To sum
$S = \sum_{k=0}^{N-1} X^{2k+1} / (2k+1)$, we simply
replace the subtractions with additions in Algorithm~1
and skip lines \ref{lst:line:spluse} and \ref{lst:line:sminusd}.

For the exp series
$S = \sum_{k=0}^{N-1} X^k / k!$, we use
different tables $u$ and $v$.
For $k! < 2^B - 1$, $u_k / v_k = 1/k!$ 
and for larger $k$, $u_k / v_k$
equals $1/k!$ times the product of all $v_i$ with
$i < k$ and distinct from $v_k$.
The $k$ such that $v_k \ne v_{k+1}$ are
$$12, 19, 26, \ldots, 264, 267, \ldots (\text{32-bit})$$ 
and
$$20, 33, 45, \ldots, 266, 273, \ldots (\text{64-bit}).$$
Algorithm~\ref{alg:atan} is modified
by skipping line \ref{lst:line:suse1} (in the next
line, the division has one less limb).
The remaining changes are that line \ref{lst:line:finalmul}
is removed, line \ref{lst:line:xsquare1} becomes $T_1 \gets X$,
and the output has $n + 1$ limbs instead of $n$ limbs.

For the sine and cosine
$S_1 = \sum_{k=0}^{N-1} (-1)^k X^{2k+1} / (2k+1)!$
and
$S_2 = \sum_{k=0}^{N-1} (-1)^k X^{2k} / (2k)!,$
we use the same $u_k, v_k$
as for exp, and skip line \ref{lst:line:suse1}.
As in the atan series, the table of powers starts with the square of $X$,
and we multiply the sine by $X$ in the end.
The alternating signs are
handled the same way as for atan, except that
line~\ref{lst:line:sminusd} becomes ${S \gets S - 1}$.
To compute sin and cos simultaneously,
we execute the main loop of the algorithm twice:
once for the sine (odd-index coefficients) and once for
the cosine (even-index coefficients), recycling the table $T$.

When computing sin and cos above circa 300 bits and exp above circa
800 bits, we optimize by just evaluating the Taylor series
for sin or sinh, after which we use $\cos(x) = \sqrt{1 - [\sin(x)]^2}$ or
$\exp(x) = \sinh(x) + \sqrt{1 + [\sinh(x)]^2}$. This removes
half of the Taylor series terms, but only saves
time at high precision due to the square root. The cosine is computed
from the sine and not vice versa to avoid the ill-conditioning
of the square root near 0.

\section{Top-level algorithm and error bounds}
\label{sect:toplevel}

Our input to an elementary function $f$
is an arbitrary-precision
floating-point number $x$ and a precision $p \ge 2$.
We output a pair of floating-point numbers $(y,z)$
such that $f(x) \in [y-z, y+z]$.
The intermediate calculations use fixed-point arithmetic.
Naturally, floating-point manipulations are used
for extremely large or small
input or output. For example,
the evaluation of
$\exp(x) = \exp(t) 2^m$, where $m$ is chosen so that
$t = x - m \log(2) \in [0, \log(2))$,
uses fixed-point arithmetic to approximate $\exp(t) \in [1,2)$.
The final output is scaled by $2^m$ after converting it to floating-point
form.

Algorithm~\ref{alg:atantop} gives pseudocode for
$\atan(x)$, with minor simplifications
compared to the actual implementation.
In reality, the quantities $(y, z)$ are not returned exactly as printed;
upon returning, $y$ is rounded to a $p$-bit floating-point number and the
rounding error of this operation
is added to~$z$ which itself is rounded up to
a low-precision floating-point number.

The variables $X, Y$ are fixed-point numbers and $Z$ is an
error bound measured in ulps.
We write $+\varepsilon$ to indicate that a result
is truncated to an $n$-limb fixed-point number, adding at most $1~\text{ulp} = 2^{-Bn}$ error
where $B = 32$ or~$64$.

After taking care of special cases, $|x|$ or $1/|x|$ is
rounded to a fixed-point number $0 \le X < 1$.
Up to two argument transformations are then applied to $X$. The
first ensures $0 \le X < 2^{-r_1}$ and the second ensures
$0 \le X < 2^{-r_1-r_2}$.
After line \ref{lst:line:lastxred}, we have (if $|x| < 1$)
$$|\atan(x)| = \atan\!\left(\frac{p_1}{2^{r_1}}\right) + \atan\!\left(\frac{p_2}{2^{r_1+r_2}}\right) + \atan(X) + \delta$$
or (if $|x| > 1$)
$$|\atan(x)| = \frac{\pi}{2} - \atan\!\left(\frac{p_1}{2^{r_1}}\right) - \atan\!\left(\frac{p_2}{2^{r_1+r_2}}\right) - \atan(X) + \delta$$
for some $|\delta| \le Z$.
The bound on $\delta$ is easily proved
by repeated application of the fact that $|\atan(t+\varepsilon)-\atan(t)| \le |\varepsilon|$
for all $t, \varepsilon \in \mathbb{R}$.

The value of $\atan(X)$ is approximated using a Taylor series.
By counting leading zero bits in $X$, we find the optimal integer $r$
with $r_1 + r_2 \le r \le Bn$
such that $X < 2^{-r}$ (we could take $r = r_1 + r_2$, but
choosing $r$ optimally is better when $x$ is tiny).
The tail of the Taylor series satisfies
$$\left|\atan(X) - \sum_{k=0}^{N-1} \frac{(-1)^k}{2k+1} X^{2k+1}\right| \le X^{2N+1},$$
and we choose $N$ such that $X^{2N+1} < 2^{-r(2N+1)} \le 2^{-w}$ where $w$
is the working precision in bits.

Values of $\atan(p_1 2^{-r_1})$, $\atan(p_2 2^{-r_1-r_2})$ and $\pi/2$
are finally read from tables with at most 1~ulp error each,
and all terms are added.
The output error bound $z$ is the sum of the Taylor series truncation
error bound and the bounds for all fixed-point rounding errors.
It is clear that
$z \le 10 \times 2^{-w}$ where the $w$ is the
working precision in bits, and that the choice of $w$
implies that $y$ is accurate to $p$ bits.
The working precision has to be increased for small input,
but the algorithm never slows down significantly since
very small input results in only a few terms of the Taylor
series being necessary.

\begin{algorithm}
  \caption{Top-level algorithm for atan}
  \label{alg:atantop}
  \begin{algorithmic}[1]
    \Require $x \not\in \{0, \pm \infty, \text{NaN}\}$ with sign $\sigma$ and exponent $e$ such that $2^{e-1} \le \sigma x < 2^e$, and a precision $p \ge 2$
    \Ensure A pair $(y,z)$ such that $\atan(x) \in [y-z, y+z]$
    \If {$e < -p/2-2$}
        \State \Return $(x, \pm 2^{3e})$ \Comment $\atan(x) = x + O(x^3)$
    \EndIf
    \If {$e > p+2$}
        \State \Return $(\sigma \pi/2, 2^{1-e})$  \Comment$\atan(x) = \pm \pi/2 + O(1/x)$
    \EndIf
    \If{$|x| = 1$}
        \State \Return $(\sigma \pi/4, 0)$
    \EndIf
    \State $w \gets p - \min(0, e) + 4$ \Comment{Working precision in bits}
    \If {$w > 4608$}
        \State \Return Enclosure for $\atan(x)$ using fallback algorithm
    \EndIf
    \State $n \gets \lceil w / B \rceil$ \Comment{Working precision in limbs}
    \If{$|x < 1$}
        \State $X \gets |x| + \varepsilon$, $Z \gets 1$
    \Else
        \State $X \gets 1 / |x| + \varepsilon$, $Z \gets 1$
    \EndIf
    \State \textbf{If} $w \le 512$ \textbf{then} $(r_1, r_2) \gets (8, 0)$ \textbf{else} $(r_1, r_2) \gets (5, 5)$
    \State $p_1 \gets \lfloor 2^{r_1} X \rfloor$ \Comment{First argument reduction}
    \If{$p_1 \ne 0$}
        \State $X \gets (2^{r_1} X - p_1)/(2^{r_1} + p_1 X) + \varepsilon, Z \gets Z + 1$
    \EndIf
    \State $p_2 \gets \lfloor 2^{r_2} X \rfloor$ \Comment{Second argument reduction}
    \If {$p_2 \ne 0$}
        \State $X \gets (2^{r_1+r_2} \!X\!-\!p_2)/(2^{r_1+r_2}\!+\!p_2 X) + \varepsilon, Z \gets Z + 1$ \label{lst:line:lastxred}
    \EndIf
    \State Compute $r_1 + r_2 \le r \le B n$ such that $0 \le X < 2^{-r}$
    \State $N \gets \lceil (w - r) / (2r) \rceil$
    \If {$N \le 2$}
        \State $Y \gets \sum_{k=0}^{N-1} \tfrac{(-1)^k}{2k+1} X^{2k+1} + 3 \varepsilon$ \Comment Direct evaluation
        \State $Z \gets Z + 3$
    \Else
        \State $Y \gets \sum_{k=0}^{N-1} \tfrac{(-1)^k}{2k+1} X^{2k+1} + 2 \varepsilon$ \Comment Call Algorithm~\ref{alg:atan}
        \State $Z \gets Z + 2$
    \EndIf
    \If{$p_1 \ne 0$} \Comment{First table lookup}
        \State $Y \gets Y + (\atan(p_1 2^{-r_1}) + \varepsilon), \; Z \gets Z + 1$
    \EndIf
    \If{$p_2 \ne 0$} \Comment{Second table lookup}
        \State $Y \gets Y + (\atan(p_2 2^{-r_1-r_2}) + \varepsilon), \; Z \gets Z + 1$
    \EndIf
    \If{$x > 1$}
        \State $Y \gets (\pi/2 + \varepsilon) - Y, Z \gets Z + 1$
    \EndIf
    \State \Return $(\sigma Y,\, 2^{-r(2N+1)} + Z 2^{-Bn})$
  \end{algorithmic}
\end{algorithm}

The code for exp, log, sin and cos implements the respective
argument reduction formulas analogously.
We do not reproduce the calculations here due to space constraints.
The reader may refer to the source code \cite{fjarbsource} for details.

Our software~\cite{Johansson:2014:ACL:2576802.2576828} chooses guard bits
to achieve $p$-bit relative accuracy with at most 1-2 ulp error in general,
but does not guarantee correct rounding, and allows the output
to have less accuracy in special cases.
In particular, sin and cos are computed to an
absolute (not relative) tolerance of $2^{-p}$ for large input, and
thus lose accuracy near the roots.
These are reasonable compromises for variable-precision interval arithmetic,
where we only require a correct enclosure of the result and
have the option to restart with higher precision if the output
is unsatisfactory.

Correct rounding (or any other strict precision policy) can be achieved
with Ziv's strategy: if the output interval $[y-z,y+z]$ does not allow
determining the correctly rounded $p$-bit floating-point approximation,
the computation is restarted with more guard bits.
Instead of starting with, say, 4 guard bits to compensate for
internal rounding error in the algorithm,
we might start with $4 + 10$ guard bits for a $2^{-10}$ probability
of having to restart.
On average, this only results in a slight increase in running time,
although worst cases necessarily
become much slower. 

\section{Benchmarks}
\label{sect:bench}

Table \ref{tab:benchmarktime} shows benchmark results done on
an Intel i7-2600S CPU running x86\_64 Linux.
Our code is built against MPIR~2.6.0.
All measurements were obtained by evaluating the function
in a loop running for at least 0.1~s and taking the best
average time out of three such runs.

The input to each function is a floating-point number close to $\sqrt{2}+1$,
which is representative for our implementation
since it involves the slowest argument reduction
path in all functions for moderate input (for input
larger than about $2^{64}$, exp, sin and cos become marginally slower
since higher precision has to be used
for accurate division by $\log(2)$ or $\pi/4$).

We include timings for the double-precision functions
provided by the default libm installed on the same system (EGLIBC~2.15).
Table \ref{tab:benchmarkratio} shows the speedup compared to MPFR~3.1.2
at each level of precision.

\begin{table}[ht!]
\centering
\caption{Timings of our implementation in microseconds. Top row: time of libm.}
\begin{tabular}{ r | r r r r r }
Bits & exp & sin & cos & log & atan \\ \hline
53 & 0.045 & 0.056 & 0.058 & 0.061 & 0.072 \\
\hline
32 & 0.26 & 0.35 & 0.35 & 0.21 & 0.20 \\
53 & 0.27 & 0.39 & 0.38 & 0.26 & 0.30 \\
64 & 0.33 & 0.47 & 0.47 & 0.30 & 0.34 \\
128 & 0.48 & 0.59 & 0.59 & 0.42 & 0.47 \\
256 & 0.83 & 1.05 & 1.08 & 0.66 & 0.73 \\
512 & 2.06 & 2.88 & 2.76 & 1.69 & 2.20 \\
1024 & 6.79 & 7.92 & 7.84 & 5.84 & 6.97 \\
2048 & 22.70 & 25.50 & 25.60 & 22.80 & 25.90 \\
4096 & 82.90 & 97.00 & 98.00 & 99.00 & 104.00 \\
\end{tabular}
\label{tab:benchmarktime}
\end{table}

\begin{table}[ht!]
\centering
\caption{Speedup vs MPFR~3.1.2.}
\begin{tabular}{ r | r r r r r }
Bits & exp & sin & cos & log & atan \\ \hline
32 & 7.9 & 8.2 & 3.6 & 11.8 & 29.7 \\
53 & 9.1 & 8.2 & 3.9 & 10.9 & 25.9 \\
64 & 7.6 & 6.9 & 3.2 & 9.3 & 23.7 \\
128 & 6.9 & 6.9 & 3.6 & 10.4 & 30.6 \\
256 & 5.6 & 5.4 & 2.9 & 10.7 & 31.3 \\
512 & 3.7 & 3.2 & 2.1 & 6.9 & 14.5 \\
1024 & 2.7 & 2.2 & 1.8 & 3.6 & 8.8 \\
2048 & 1.9 & 1.6 & 1.4 & 2.0 & 4.9 \\
4096 & 1.7 & 1.5 & 1.3 & 1.3 & 3.1 \\
\end{tabular}
\label{tab:benchmarkratio}
\end{table}

\begin{table}[ht!]
\centering
\caption{Top rows: timings in microseconds for
quadruple (113-bit) precision, except QD which gives 106-bit precision.
Bottom rows: timings for quad-double (212-bit) precision.
Measured on an Intel T4400 CPU.}
\begin{tabular}{ l | r r r r r }
               & exp  & sin  & cos  & log  & atan \\ \hline
MPFR           & 5.76 & 7.29 & 3.42 & 8.01 & 21.30 \\
libquadmath    & 4.51 & 4.71 & 4.57 & 5.39 & 4.32 \\
QD (dd)        & 0.73 & 0.69 & 0.69 & 0.82 & 1.08 \\
Our work       & 0.65 & 0.81 & 0.79 & 0.61 & 0.68  \\ \hline
MPFR & 7.87 & 9.23 & 5.06 & 12.60 & 33.00 \\
QD (qd)        & 6.09 & 5.77 & 5.76 & 20.10 & 24.90 \\
Our work & 1.29 & 1.49 & 1.49 & 1.26 & 1.23  \\
\end{tabular}
\label{tab:benchmarkquad}
\end{table}

Table \ref{tab:benchmarkquad} provides a comparison at IEEE 754 quadruple (113-bit)
precision against MPFR and the libquadmath included with GCC 4.6.4. We include timings for
the comparable double-double (``dd'', 106-bit) functions provided by version 2.3.15 of
the QD library~\cite{hida2007library}.
Table~\ref{tab:benchmarkquad} also compares
performance at quad-double (``qd'', 212-bit) precision against MPFR and QD.
The timings in Table~\ref{tab:benchmarkquad} were obtained on a
slower CPU than the
timings in Table~\ref{tab:benchmarktime}, which we used due to the GCC version
installed on the faster system
being too old to ship with libquadmath.

At low precision, a function evaluation
with our implementation takes less than half a microsecond,
and we come within an order of magnitude of the default libm at 53-bit precision.
Our implementation holds up well around 100-200 bits of precision,
even compared to a library specifically designed for this range (QD).

Our implementation is consistently faster than MPFR.
The smallest speedup is achieved for the cos function, as the argument reduction without table lookup
is relatively efficient
and MPFR does not have to evaluate the Taylor series for
both sin and cos. The speedup is largest for atan, since
MPFR only implements the bit-burst algorithm for this function,
which is ideal only for very high precision.
Beyond 4096 bits, the asymptotically fast algorithms implemented
in MPFR start to become competitive for all functions,
making the idea of using
larger lookup tables to cover even higher precision
somewhat less attractive.

Differences in accuracy should be considered when
benchmarking numerical software.
The default libm, libquadmath, and QD do not provide error bounds.
MPFR provides the strongest guarantees (correct rounding).
Our implementation provides rigorous error bounds, but allows
the output to be less precise than correctly rounded.
The 20\% worse speed at 64-bit precision compared
to 53-bit precision gives an indication of the overhead that would
be introduced by providing correct rounding
(at higher precision, this factor would be smaller).

\section{Future improvements}

Our work helps reduce the performance gap between double
and multiple precision.
Nonetheless, our approach is not
optimal at precisions as low as 1-2 limbs,
where rectangular splitting has no advantage over
evaluating minimax polynomials with Horner's rule,
as is generally done in libraries targeting
a fixed precision.

At very low precision, GMP functions are likely inferior to
inlined double-double and quad-double arithmetic or similar,
especially if the floating-point operations are vectorized.
Interesting alternatives designed to exploit hardware parallelism
include the carry-save library used for
double-precision elementary functions with correct rounding in CR-LIBM~\cite{daramy2003cr,defour2003},
the recent SIMD-based multiprecision
code \cite{van2014modular}, and implementations targeting GPUs~\cite{thall2006extended}.
We encourage further comparison of these options.

Other improvements are possible at higher precision.
We do not need to compute every term to a precision
of $n$ limbs in Algorithm~\ref{alg:atan} as the contribution
of term~$k$ to the final sum is small when $k$ is large.
The precision should rather be changed progressively.
Moreover, instead of computing an $(n \times n)$-limb fixed-point product
by multiplying exactly and throwing away the low $n$ limbs, we
could compute an approximation of the high part in about
half the time (unfortunately, GMP does not
currently provide such a function).

Our implementation of the elementary functions outputs
a guaranteed error bound whose proof of correctness
depends on a complete error analysis done by hand,
aided by some exhaustive computations.

To rule out any superficial bugs, we have tested the code
by comparing millions of
random values against MPFR.
We also test the code against itself for millions of random inputs
by comparing the output at different levels of precisions or at
different points connected by a functional equation.
Random inputs are generated non-uniformly to increase
the chance of hitting corner cases.
The functions are also tested indirectly by being used
internally in many higher transcendental functions.

Nevertheless, since testing cannot completely rule out human error,
a formally verified implementation would be desirable.
We believe that such a proof is feasible.
The square root function in GMP is implemented at a similar level
of abstraction, and it has been proved correct
formally using Coq~\cite{bertot2002proof}.

\section*{Acknowledgments}

This research was partially funded by ERC Starting Grant ANTICS 278537.
The author thanks the anonymous referees for valuable feedback.

\bibliographystyle{plain}
\bibliography{references.bib}

\end{document}